\documentclass[10pt,twocolumn,aps,pra,superscriptaddress,longbibliography,floatfix]{revtex4-2}

\usepackage{graphicx}
\usepackage{xcolor}
\usepackage{dsfont}
\usepackage{amsmath,amssymb,amsthm}
\usepackage{tikz}
\usepackage[normalem]{ulem}
\usepackage{hyperref}
\usepackage[capitalise]{cleveref}
\usepackage{float}
\usepackage{enumitem}
\makeatletter
\let\newfloat\newfloat@ltx
\let\float@end\float@end@ltx
\makeatother
\usepackage{algorithm}
\makeatletter
\renewcommand\fnum@algorithm{\fname@algorithm~\thealgorithm}
\makeatother
\usepackage{algpseudocode}
\algrenewcommand\algorithmicrequire{\textbf{Input:}}
\algrenewcommand\algorithmicensure{\textbf{Output:}}
\hypersetup{colorlinks=true,linkcolor=blue,citecolor=blue,urlcolor=blue,
 pdftitle={Multi-particle quantum walk search on a two-dimensional lattice},
 pdfauthor={Goncalo Meneses and D. O. Oriekhov}}
\graphicspath{{figures/}}
\allowdisplaybreaks[2]
\newcommand{\ket}[1]{\lvert#1\rangle}
\newcommand{\bra}[1]{\langle#1\rvert}
\newcommand{\ketbra}[2]{\ket{#1}\!\bra{#2}}

\newcommand{\Tr}{\operatorname{Tr}}

\begin{document}
\title{Multi-particle quantum walk search on a two-dimensional lattice}
\date{\today}
\author{Gon\c{c}alo Meneses}
\affiliation{Kavli Institute of Nanoscience, Delft University of Technology, 2628 CJ Delft, the Netherlands}
\author{D. O. Oriekhov}
\affiliation{Kavli Institute of Nanoscience, Delft University of Technology, 2628 CJ Delft, the Netherlands}

\begin{abstract}
In the present paper we study the formulation of a search problem and the efficiency of multi-particle quantum walk search algorithms. The discrete-time evolution of each particle in space is treated according to Grover's quantum walk evolution operator. We focus on the local interactions models, where the particles' states are modified only when two or more of them appear on the same grid site. The main focus is given to the two-dimensional grid in space. We study the dependence of the search algorithm asymptotic success time and probability on the interaction strength and the generated maximal entanglement between particles. We show that the interacting and non-interacting two- and three-particle search algorithm attains a constant factor improvement over the single-particle counterpart. In addition, we provide a new generalized model of interactions for multi-particle quantum walks.

\end{abstract}

\maketitle

\section{Introduction}
\label{sec:introduction}

Quantum walks provide a framework for quantum computation in which
information is encoded in positional and internal degrees of freedom of a particle
and processed through their coherent evolution on a graph. It was shown by a number of studies \cite{Childs2009,LoCooEve2010,SinghChawSar2021} that quantum walk algorithms are universal models of quantum computation. In discrete-time constructions, this is achieved by establishing a proper mapping between the gate-based operations and the evolution operator of a quantum walk. At the same time, a single quantum walk step can implement a structured evolution equivalent to a sequence of one- and two-qubit gates. Quantum walks thus hold the potential to fully implement gate-based quantum algorithms or to replace subroutines within these. A notable example of the latter is the Childs-Kothari-Somma (CKS) algorithm for solving sparse systems of linear equations, whose quantum-walk-based construction exponentially improved the runtime's dependence on the target precision compared with HHL~\cite{ChildsKoSom2017}.

Quantum walk algorithms support a natural way of implementation: one particle is loaded into a quantum simulator with several available positions or virtual states and then evolves according to its Hamiltonian. Quantum walks have been experimentally realized on a variety of quantum computing and quantum simulation platforms, including nuclear magnetic resonance processors~\cite{RyLaBoi2005}, trapped ions~\cite{SchmitzMatSchnei2009,ZahKirchGer2010}, optically trapped neutral atoms~\cite{KarForChoi2009}, photonic systems~\cite{BroomeFeLan2010,SchreiCasPo2010}, ultracold atoms supporting momentum-space walks~\cite{DaGreschGroi2019} and semiconductor quantum dot spin arrays~\cite{FaJiZhang2025}. Implementations on arbitrary spatial networks have also been proposed with Rydberg atom arrays, achieving a quadratic speedup in spatial search~\cite{AlSanOmar2026}.
The evolution of a particle within such a simulator already represents a quantum walk algorithm, even without additional external control pulses. As the most straightforward example, the evolution under the same unitary operator could realize the quantum walk search algorithm \cite{ShenKemWha2003,AmKemRi2004}. Apart from applying artificial control pulses, it is also possible to load several particles into the same quantum simulator. This latter scenario generously extends the total size of the Hilbert space available for the computation and adds another natural and controllable parameter to an evolution operator of a system, the particle number. In this paper, we focus on the question of how a multi-particle quantum walk implements a quantum search algorithm and how its efficiency compares to that of the single-particle quantum walk search.

To address this question, we first review the efficiency of known quantum walk search algorithms.
The first references reported a quantum walk search algorithm back in the early 2000s \cite{ShenKemWha2003,AmKemRi2004}. The problem was formulated as follows: suppose we have a regular grid of points in space and the quantum particle can move and change its spin state according to a set of local rules. Then, in one node of this grid, they are changed. What is the probability that the particle will localize at that modified (target) node and how many time steps would this localization require? The answer was shown to depend strongly on the dimensionality of the grid. For a database of size $N$, it was found \cite{AmKemRi2004} that there is no quantum advantage in a 1D grid, the asymptotic search time is $T=O(\sqrt{N}\log N)$ in 2D with a respective localization probability of $O(1/\log N)$ and finally, in higher dimensions, one finds an asymptotic time of $T=O(\sqrt{N})$, as in Grover's algorithm.
In addition, it is possible to further improve the time scaling to $T=O(\sqrt{N \log N})$ by using an ancilla qubit in the 2D grid,
which is implemented in Tulsi's algorithm \cite{Tul2008}. Below, for each particle, we select the same type of quantum walk rules in a 2D grid as in Refs.~\cite{AmKemRi2004,RoKaDiMol2023}.

The second important part in our study is the interaction model. First, we note that the vast majority of modern quantum simulators efficiently suppress long-range cross-talk between sites \cite{SchmitzMatSchnei2009,ZahKirchGer2010,KarForChoi2009,BroomeFeLan2010,SchreiCasPo2010,FaJiZhang2025}. In the experiment, this suppression is made and tuned to achieve the high coherence of local gate operations and readout procedures.
For our paper, it is thus well motivated to study the local interactions. The locality means that the particles' states are modified by an interaction operator only if two or more of them appear at the same site at a given discrete time step. Multiple local interaction models were already studied in the literature on quantum walks \cite{CosMePor2019,MitSo2026}. For the study in this paper, we select the so-called phase interaction, which changes the phase of the wave function if, at least, two particles appear on the same site and the hard-collision interaction, which implies that particles move in different directions after meeting at the same spatial vertex of the grid, according to a set of local rules.

As the final ingredient in the studied question, we should list the existing tunable hyperparameters in the system. All of them could be optimized to improve the algorithm's performance. In the simplest case, the optimization will influence the constant prefactors before the success time or search probability asymptotics. These hyperparameters are the marked site operator and the interaction strength for certain interaction models. The rigorous theory behind the optimal selection of these operators is not well developed even for single-particle quantum walks \cite{AmKemRi2004,RoKaDiMol2023} and the most natural way is to explore this problem numerically.
To evaluate the role of the second parameter, the interaction strength, we develop a generalized model of local interactions that smoothly varies between non-interacting and maximally interacting particles. We implement this for the collision model \cite{MitSo2026}.

The paper is organized as follows. In Sec.~\ref{sec:quantum-search-problem}, we formulate the quantum search problem for a multi-particle quantum walk and define which joint configurations count as finding the marked vertex. Section~\ref{sec:model} constructs the search algorithm and the general interaction operator for $P$ walkers and then develops the phase and HPP interaction models. Section~\ref{sec:halo_exchange} presents the distributed halo exchange algorithm we use to simulate the walk on large lattices. In Sec.~\ref{sec:search_results}, we establish localization and inter-particle entanglement in the non-interacting search and, in Sec.~\ref{sec:interaction_results}, we study how HPP collisions and uniform phase interactions modify this behavior. Section~\ref{sec:scaling} examines how the runtime and success probability scale with the database size. We discuss the efficiency of the multi-particle quantum walk algorithm, our numerical findings and the future outlook in Sec.~\ref{sec:discussion}.

\section{Abstract formulation of a multi-particle search problem}
\label{sec:quantum-search-problem}

Consider finding a single marked vertex $\mathbf m$ on a periodic
$d$-dimensional grid with $N=L^d$ vertices, using a quantum walk that moves
only between adjacent vertices. For the unstructured search problem the Grover's
algorithm needs $\Theta(\sqrt N)$ queries~\cite{Gro1996}. Coined quantum
walks match this bound for $d\geq3$, but in two dimensions the algorithm of
Ambainis, Kempe and Rivosh (AKR) succeeds only with probability
$\Theta(1/\log N)$ after $\Theta(\sqrt{N\log N})$ steps, i.e., in
$O(\sqrt N\log N)$ steps with amplitude amplification~\cite{AmKemRi2004}.
Tulsi's ancilla-controlled walk reduces this to
$O(\sqrt{N\log N})$~\cite{Tul2008} and, to our knowledge, the question of whether any local
algorithm reaches $O(\sqrt N)$ on the two-dimensional grid remains open.

The logarithm has a simple physical origin. A quantum walk finds a marked
vertex in a number of steps of order the square root of the classical
hitting time, i.e., the expected time a classical random walk on the same
graph needs to reach it~\cite{Sze2004,KroMagOzols2016}. On a
two-dimensional grid, this hitting time is $\Theta(N\log N)$ rather than
$\Theta(N)$. The reason is that the planar random walk is recurrent: it
keeps returning to regions it has already visited, so it explores the grid
inefficiently and every length scale up to the system size contributes
equally to the time needed to reach $\mathbf m$. In the quantum walk, these
contributions appear as long-wavelength modes that accumulate
logarithmically, which both slows down the localization at $\mathbf m$ and
reduces the final success probability~\cite{AmKemRi2004,Tul2008}. In three
or more dimensions, the random walk is transient, the hitting time is
$\Theta(N)$ and the overhead disappears, which makes two dimensions the
critical case~\cite{ChildsGold2004b}. Since $P$ independent classical
walkers reduce the hitting time by roughly a factor of $P$, this motivates
asking whether several quantum walkers, possibly interacting, can do
better.

\begin{figure}[tb]
\centering
\begin{tikzpicture}[x=0.52cm,y=0.52cm,font=\small]
 \fill[blue!17] (0,2) rectangle (5,3);
 \fill[blue!17] (2,0) rectangle (3,5);
 \foreach \k in {0,...,4} {
   \fill[orange!55] (\k,\k) rectangle +(1,1);
 }
 \fill[violet!60] (2,2) rectangle (3,3);
 \draw[step=1,gray!70] (0,0) grid (5,5);
 \draw[->] (0,-0.35)--(5.5,-0.35) node[right] {$\mathbf x_1$};
 \draw[->] (-0.35,0)--(-0.35,5.5) node[above] {$\mathbf x_2$};
 \node[below] at (2.5,-0.35) {$\mathbf m$};
 \node[left] at (-0.35,2.5) {$\mathbf m$};
 \fill[blue!17] (6,4) rectangle +(0.45,0.45);
 \node[right,align=left,font=\footnotesize] at (6.55,4.2) {At least one\\walker on $\mathbf m$ (i)};
 \fill[orange!55] (6,2.65) rectangle +(0.45,0.45);
 \node[right,align=left,font=\footnotesize] at (6.55,2.85) {Coincident\\positions};
 \fill[violet!60] (6,1.3) rectangle +(0.45,0.45);
 \node[right,align=left,font=\footnotesize] at (6.55,1.5) {Both walkers\\on $\mathbf m$ (ii)};
\end{tikzpicture}
\caption{Joint position space of two walkers; each axis lists lattice
vertices (five shown). Blue, including the violet intersection: target set
of problem (i), on which the oracle acts. Violet: the single target
configuration of problem (ii). An on-site interaction acts on the diagonal
(orange and violet). On a two-dimensional lattice each axis runs over all
$N$ vertices.}
\label{fig:model}
\end{figure}
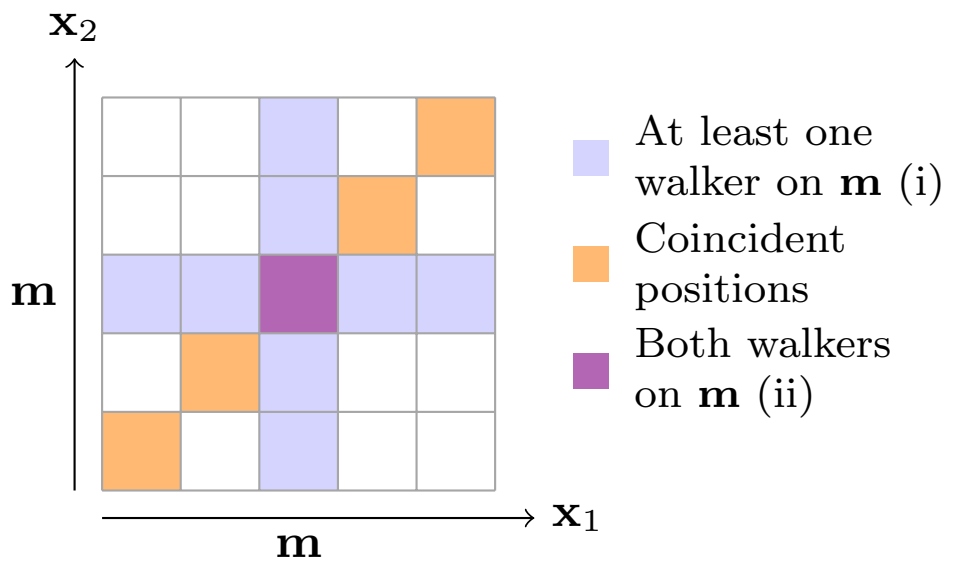

In single-particle search, the database is the set of lattice vertices. The coin
conditions the motion of the walker but carries no database information, so
a measurement succeeds whenever the walker is found at marked vertex $\mathbf m$,
whatever its coin state~\cite{ShenKemWha2003,AmKemRi2004}. With $P$ walkers
on the same grid, the joint position space has $N^P$ configurations and
many of them contain a walker at $\mathbf m$. The solution therefore no
longer corresponds to a single configuration and one must specify which
configurations count as finding it. Two criteria are straightforward:
\begin{enumerate}
  \item[(i)] \emph{any-walker criterion}: the search succeeds if at least
  one walker is found at $\mathbf m$, irrespective of the positions of the
  others;
  \item[(ii)] \emph{all-walker criterion}: the search succeeds only if all
  walkers are found at $\mathbf m$ simultaneously.
\end{enumerate}
Both can be checked by counting the walkers at the marked vertex and
Fig.~\ref{fig:model} shows the corresponding configurations for $P=2$. The
two criteria pose different problems. Criterion (ii) singles out one
configuration among $N^P$, so it is a search over the full configuration
space rather than over the lattice. Moreover, interactions that suppress
multiple occupancy, in analogy with the Pauli principle, forbid it
altogether. Criterion (i) instead asks the spatial question of where
$\mathbf m$ lies. Its solution set contains a fraction
$1-(1-1/N)^P\simeq P/N$ of all configurations for $N\gg P$ and finding a
single walker at $\mathbf m$ answers it. We adopt the any-walker criterion
throughout. This criterion fixes the marked subspace of the joint Hilbert
space and thereby how the search operator must act on the joint state (see Sec.~\ref{sec:model}). This setting also differs
from search with several marked vertices, where the arrangement of the
marked vertices affects performance~\cite{RoKaDiMol2023}. Here, several
walkers instead search collectively for a single marked vertex.
 
\section{Mathematical models for multi-particle quantum walks}
\label{sec:model}

In this section, we begin by defining the single-particle Grover walk from Ambainis, Kempe and Rivosh (AKR)~\cite{AmKemRi2004}, followed by the construction of the joint marked coin needed to generalize their algorithm to the multi-particle domain. This many-body form of the coin operator follows from the construction we outlined in the previous section, i.e., the any-walker convention for finding the solution to the search problem. Once we have formulated our multi-particle quantum random walk algorithm, we extend the two-particle interaction framework of Mittal and Sowi\'{n}ski~\cite{MitSo2026}, including their phase interaction model. Furthermore, we introduce a coin-flipping HPP interaction from Costa, Melo and Portugal~\cite{CosMePor2019}.



\subsection{Single-particle quantum walk search algorithm}

We consider an $L\times L$ square lattice with periodic boundary conditions, where each vertex encodes one of the $N=L^2$ database elements. One of these elements is then the solution to the search problem, or the marked element $\ket{\mathbf{m}}$. All other elements are represented by $\ket{\mathbf{x}}\neq\ket{\mathbf{m}}$. Here, periodic boundaries are introduced in order to study the localization of the wavefunction around the marked vertex without introducing edge effects. The Hilbert space of a walker $a$ then consists of a four-dimensional coin subspace and an $N$-dimensional position subspace,
\begin{equation}
 \mathcal H^{(a)}=\mathcal H_c^{(a)}\otimes\mathcal H_p^{(a)}.
\end{equation}

The four dimensions of the coin basis then represent the cardinal directions on the grid, $\ket{\mathbf d}\in\{\ket{\uparrow},\ket{\rightarrow},\ket{\leftarrow},\ket{\downarrow}\}$. Following, as promised, the AKR~\cite{AmKemRi2004} single-particle quantum walk search construction, we use the Grover coin
\begin{equation}
 G=2\ketbra{s_c}{s_c}-I_c,
 \qquad \ket{s_c}=\frac12\sum_{\mathbf d}\ket{\mathbf d},
 \label{eq:grover_coin}
\end{equation}
and the flip-flop shift
\begin{equation}
 S_{\mathrm{ff}}\ket{\mathbf d,\mathbf x}
 =\ket{\bar{\mathbf{d}},\mathbf x+\mathbf d},
 \label{eq:flipflop_shift}
\end{equation}
where position addition is modulo $L$, due to the periodic boundary conditions. In a discrete-time coined quantum walk, the dynamics of the system are governed by the discrete application of these two operators, specifically by first applying the coin operator and immediately afterwards the shift operator. This means that the overall evolution operator of the quantum walk is
\begin{equation}
    U=S_{\mathrm{ff}}(G\otimes I_p).
\end{equation}

However, this is only the definition of a general quantum walk. In order for the dynamics to allow for localization of the particle at a target node, we must be able to differentiate this node from the other vertices on the grid. To do this, we define the marked subspace projector $P_m=\ketbra{\mathbf m}{\mathbf m}$. Using this operator and a similar approach to the one Grover's algorithm applies in the circuit model, we apply a relative phase of $\pi$ at this specific element. We do this by introducing a perturbation to the coin operator:
\begin{equation}
    C'=G\otimes(I_p-P_m)-I_c\otimes P_m.
    \label{eq:single_marked_coin}
\end{equation}
The final evolution operator used to perform a single-particle quantum walk search algorithm is then
\begin{equation}
    U'=S_{\mathrm{ff}}\,C'.
    \label{eq:single_search_evol}
\end{equation}
Combining these ingredients, the full single-particle quantum walk spatial search algorithm reads:
\begin{enumerate}[label=\textbf{Step \arabic*:}, leftmargin=*, itemsep=2pt]
    \item Prepare the overall state of the system in the uniform superposition $\ket{\psi_0}=\ket{s_c}\otimes\frac{1}{\sqrt N}\sum_{\mathbf x}\ket{\mathbf x}$.
    \item Apply the evolution operator $U'$ a total of $T=O\left(\sqrt{N\log N}\right)$ steps.
    \item Measure the position subspace and verify that the obtained state is $\ket{\mathbf m}$.
    \item If the result is not $\ket{\mathbf m}$ repeat steps 1 to 3.
\end{enumerate}

\subsection{Moving to the multi-particle domain}

For multiple walkers, the most natural way to extend the problem, in a similar fashion to Mittal and Sowi\'{n}ski~\cite{MitSo2026}, is to take the Hilbert space of $P$ particles to be a tensor product of all individual particles' Hilbert spaces,
\begin{equation}
    \mathcal H=\bigotimes_{a=1}^P\mathcal H^{(a)}.
\end{equation}
A first attempt to build a multi-particle quantum walk search is then to evolve each walker with its own single-particle perturbed evolution operator (Eq.~\ref{eq:single_search_evol}),
\begin{equation}
    \mathcal{U}_{\otimes} = U'^{\otimes P}.
\end{equation}
This approach has two shortcomings. First, it applies the perturbed coin $C'$ to each particle's subspace independently. Any increase in success probability obtained by measuring the output of this walk is therefore equivalent to running $P$ independent single-particle searches in parallel, which offers no benefit beyond repeating a single search $P$ times. Our goal is instead to leverage quantum effects, be it superposition, coupling, or some type of interaction, to achieve an improved speedup. Second and more fundamentally, independent marked coins do not implement the any-walker criterion. This already becomes visible for $P=2$. If we take the joint position state $\ket{\mathbf m, \mathbf m}$, where both particles coincide at the marked element, the two minus signs cancel,
\begin{equation}
 C'^{\otimes2}
 \ket{s_c,s_c}\ket{\mathbf m,\mathbf m}
 =+\ket{s_c,s_c}\ket{\mathbf m,\mathbf m},
\end{equation}
making the $\ket{\mathbf{m}, \mathbf{m}}$ target state unmarked. More generally, a configuration containing $k$ marked walkers acquires the factor $(-1)^k$. Independent marked coins thus distinguish between even and odd target occupancies, whereas our success criterion treats every nonzero occupancy in the same way, i.e., all the cases where at least one walker is at the marked vertex.

To achieve a joint search consistent with the any-walker criterion, the coin must give the same $\pi$ phase shift to every position state where at least one walker is at the marked element. Instead of a tensor product of single-particle coins, we therefore design a joint multi-particle coin operator. This requires generalizing $P_m$ to project not onto the individual marked subspace of each particle, but onto the subspace of $\mathcal{H}$ where at least one of the walkers is at the marked element. The easiest way to construct this projector is through the complementary event, where every walker avoids $\ket{\mathbf m}$, with the projector
\begin{equation}
    \left(I_p - P_m\right)^{\otimes P}.
\end{equation}
The generalization of $P_m$ is then
\begin{equation}
    \mathcal{P}_{m} = I_p^{\otimes P} - \left(I_p - P_m\right)^{\otimes P}.
\end{equation}
By construction, $\mathcal P_m$ acts as the identity on the entire marked subspace, independently of how many walkers occupy $\mathbf m$. For two particles, for example, $\ket{\mathbf m, \mathbf x}$, $\ket{\mathbf x, \mathbf m}$ and $\ket{\mathbf m, \mathbf m}$ are all eigenstates of $\mathcal{P}_m$ with eigenvalue $+1$. Any phase applied through $\mathcal P_m$ is therefore shared equally by all marked configurations, which removes the parity problem of the naive construction. Choosing this phase to be $\pi$, as in the single-particle case, our multi-particle perturbed coin is
\begin{equation}
    \mathcal{C}'=G^{\otimes P}\otimes(I_p^{\otimes P}-\mathcal P_m)-I_c^{\otimes P}\otimes\mathcal P_m,
    \label{eq:perturbed_multi_coin}
\end{equation}
where we have reordered the tensor factors as $\mathcal H_c^{\otimes P}\otimes\mathcal H_p^{\otimes P}$. The shift operator is still given by the individual flip-flop shift of each particle, $S_{\mathrm{ff}}^{\otimes P}$, leading to the multi-particle search evolution operator
\begin{equation}
    \mathcal{U}' = S_{\mathrm{ff}}^{\otimes P}\,\mathcal{C}'.
    \label{eq:multi_search_evol}
\end{equation}
Setting $P=1$ recovers the single-particle perturbed coin and evolution operator of Eqs.~\eqref{eq:single_marked_coin} and \eqref{eq:single_search_evol}. For $P>1$, however, the joint coin acts on the coins of all walkers whenever at least one of them occupies the marked vertex. For instance, in the state $\ket{\mathbf m, \mathbf x}$ with $\mathbf x\neq\mathbf m$, the second walker's coin receives $-I_c$ instead of $G$, even though that walker is not at the marked vertex. The evolution of each walker thus depends on the positions of the others, which couples the different walks and generates entanglement from the initial product state.

We refer to Eq.~\eqref{eq:multi_search_evol} as the non-interacting search, since no explicit interaction between the walkers has been added. Nevertheless, the walkers remain coupled through $\mathcal C'$. This search provides the reference against which we will assess the role of local interactions on unmarked vertices, as well as any additional effect of a collision operator on top of the joint search dynamics already generated by the marked coin.

Now that we have all the necessary ingredients, we can finally outline the (non-interacting) multi-particle quantum walk spatial search algorithm:
\begin{enumerate}[label=\textbf{Step \arabic*:}, ref=\arabic*, leftmargin=*, itemsep=2pt]
    \item Prepare the system in the uniform superposition $\ket{\Psi_0}=\left(\ket{s_c}\otimes\ket{s_p}\right)^{\otimes P}$, where $\ket{s_p}=\frac{1}{\sqrt N}\sum_{\mathbf x}\ket{\mathbf x}$. \label{step:multi_prepare}
    \item Apply the evolution operator $\mathcal{U}'$ for a total of $T$ steps. \label{step:multi_evolve}
    \item Measure the position registers of all walkers and check whether at least one walker is at the marked vertex $\mathbf m$. \label{step:multi_measure}
    \item If no walker is at $\mathbf m$, repeat Steps~\ref{step:multi_prepare} to \ref{step:multi_measure}.
\end{enumerate}
In contrast to the single-particle case, the optimal number $T$ of steps is not known for the multi-particle search. We will therefore determine it through numerical simulations.

\subsection{General interaction framework}
\label{subsec:interactions}

Our interaction model extends the two-walker framework of Mittal and Sowi\'{n}ski~\cite{MitSo2026} to pairwise local interactions among $P$ walkers. In their formulation, an arbitrary unitary acts on the coin degrees of freedom of two walkers when their positions coincide and acts as the identity elsewhere. To retain this locality for more walkers, each pair must be allowed to interact whenever its two walkers meet, irrespective of where the remaining $P-2$ walkers are. This requires coincidence projectors on the full joint position space.

For each pair $a<b$, we define
\begin{equation}
 \Pi_{ab}=\sum_{\mathbf{x}_1, \ldots, \mathbf{x}_P}
 \delta_{\mathbf x_a,\mathbf x_b}
 \ketbra{\mathbf{x}_1, \ldots, \mathbf{x}_P}{\mathbf{x}_1, \ldots, \mathbf{x}_P}.
 \label{eq:pair_coincidence_projector}
\end{equation}
Let $\mathcal O_{ab}$ be a unitary acting on the coins of walkers $a$ and $b$ and as the identity on all other coins. The corresponding position-conditioned interaction operator is
\begin{equation}
 V_{ab}=I+(\mathcal O_{ab}-I_c^{\otimes P})\otimes\Pi_{ab},
 \label{eq:pair_interaction}
\end{equation}
where $I = I_c^{\otimes P} \otimes I_p^{\otimes P}$. On coincident configurations, $V_{ab}$ applies $\mathcal O_{ab}$ to the coins, whereas on all other configurations the second term vanishes and $V_{ab}$ acts as the identity. Since these are complementary position subspaces and the coin operation is unitary, $V_{ab}$ is unitary.

The full interaction is the ordered product over all $P(P-1)/2$ pairs,
\begin{equation}
 \mathcal V=\prod_{a<b}V_{ab}
 =V_{12}V_{13}\cdots V_{(P-1)P},
 \label{eq:general_P_interaction}
\end{equation}
where the factors are ordered lexicographically in $(a,b)$ and the rightmost factor is applied first. This order matters whenever the pair operations do not commute. For example, for three walkers, $\mathcal O_{12}$ and $\mathcal O_{13}$ share the coin of walker 1 and reversing their order can change the evolution when all three walkers coincide. A specific interaction within this framework is therefore defined by both the pair operations $\mathcal O_{ab}$ and their ordering. For the diagonal phase interactions studied below, both $\mathcal O_{ab}$ and $\Pi_{ab}$ are diagonal, so all factors $V_{ab}$ commute. The ordering is then irrelevant and the interaction is defined for arbitrary $P$ without further choices.

This interaction operator is defined independently of the search oracle and can either be embedded in the evolution operator or applied as a separate operator after each step of the walk. In this work, we consider two different realizations: a coin-dependent phase interaction, similar to that of Mittal and Sowi\'{n}ski~\cite{MitSo2026} and a tunable Hardy--Pomeau--de Pazzis (HPP) collision introduced by Costa, Melo and Portugal~\cite{CosMePor2019}, which allows us to study coin-flipping interactions. The phase model is defined for arbitrary $P$ and simulated here for two and three walkers. The HPP model, however, is studied only for two walkers, since its pair operations do not commute when two pairs share a walker and its extension to more walkers would therefore require choosing an ordering convention. Both models have tunable parameters: for the phase interaction, these are the phases of the coin channels, as we will see in the next section, while for the HPP collision, it is the interaction strength. This allows us to perform a more in-depth study of the role of interactions in the search algorithm and of their potential to improve multi-particle search.

\subsection{Restriction to a phase interaction}
\label{subsec:phase_inter}

In the previous section, we mentioned that the particle-pair operator $\mathcal O_{ab}$ encodes the desired local interaction in the multi-particle quantum walk framework. Since $\mathcal O_{ab}$ is unitary, we can write it in terms of its Hamiltonian generator $H_{ab}$, i.e., $\mathcal O_{ab} = e^{iH_{ab}}$. On a two-dimensional lattice, each particle's coin subspace is four-dimensional, so the joint coin space of a pair $(a,b)$ is 16-dimensional. It is spanned by the joint direction states $\ket{\mathbf{d}_a, \mathbf{d}_b}$, with $\mathbf{d}_a, \mathbf{d}_b \in \left\{\uparrow, \rightarrow, \leftarrow, \downarrow\right\}$. Until now, we have only used these direction labels for the coin basis states of the 2D lattice; however, it is convenient to decompose each of them into a two-qubit encoding, $\ket{\mathbf{d}_i} = \ket{c_{x_i}, c_{y_i}}$, with the correspondence $\ket{\uparrow} = \ket{00}$, $\ket{\rightarrow} = \ket{01}$, $\ket{\leftarrow} = \ket{10}$ and $\ket{\downarrow} = \ket{11}$. We choose this encoding such that opposite directions are bitwise complements of each other, i.e., the direction opposite to $\mathbf{d}_i$ is $\ket{\bar{\mathbf{d}}_i} = \ket{\bar{c}_{x_i}, \bar{c}_{y_i}}$, with $\bar{c} = 1 - c$.

We now restrict our interaction model to phase interactions, similar to those of Mittal and Sowi\'{n}ski~\cite{MitSo2026}, i.e., we take $\mathcal{O}_{ab}$ and hence $H_{ab}$, to be diagonal in the joint direction basis. Physically, when walkers $a$ and $b$ share a site, the interaction then preserves their individual coin states and only assigns a phase that depends on their joint direction configuration,
\begin{equation}
    H_{ab} = \sum_{\mathbf{d}_a, \mathbf{d}_b} \lambda(\mathbf{d}_a, \mathbf{d}_b) \ketbra{\mathbf{d}_a, \mathbf{d}_b}{\mathbf{d}_a, \mathbf{d}_b},
    \label{eq:gen_phase_ham}
\end{equation}
with one real phase $\lambda(\mathbf{d}_a, \mathbf{d}_b)$ per joint direction state. Since $V_{ab}$ acts as the identity on non-coincident configurations, these phases are relative to the rest of the state, so even a uniform choice of $\lambda$ is not a global phase.

The general form in Eq.~\eqref{eq:gen_phase_ham} still contains 16 independent phases, one for each joint direction state. To reduce this parameter space, we group the joint direction states according to the relative orientation of the two walkers. Identifying each direction with its unit vector on the lattice, we distinguish
\begin{itemize}
    \item Parallel states, with $\mathbf{d}_1 = \mathbf{d}_2$;
    \item Antiparallel states, with $\mathbf{d}_1 = \bar{\mathbf{d}}_2$;
    \item Perpendicular states, with $\mathbf{d}_1 \cdot \mathbf{d}_2 = 0$.
\end{itemize}
Assigning one phase to each class, Eq.~\eqref{eq:gen_phase_ham} reduces to
\begin{equation}
    H_{ab} = \lambda_{\parallel} P_{\parallel}^{(ab)} + \lambda_{\rightleftarrows} P_{\rightleftarrows}^{(ab)} + \lambda_{\perp} P_{\perp}^{(ab)},
    \label{eq:ham_three_class}
\end{equation}
with the projectors
\begin{subequations}
\begin{align}
    P_{\parallel}^{(ab)} &= \sum_{\mathbf{d}} \ketbra{\mathbf{d}, \mathbf{d}}{\mathbf{d}, \mathbf{d}},
    \\
    P_{\rightleftarrows}^{(ab)} &= \sum_{\mathbf{d}} \ketbra{\mathbf{d}, \bar{\mathbf{d}}}{\mathbf{d}, \bar{\mathbf{d}}},
    \\
    P_{\perp}^{(ab)} &= I_c^{\otimes 2} - P_{\parallel}^{(ab)} - P_{\rightleftarrows}^{(ab)}.
\end{align}
\end{subequations}
The parallel and antiparallel subspaces are each spanned by four joint direction states, while the perpendicular subspace is spanned by eight.

This means that we have subdivided the collinear class, consisting of pairs with the same or opposite directions, into a parallel and an antiparallel subclass and one might thus ask why we do not do the same for the perpendicular class, for instance by dividing it into pairs related by a clockwise quarter-turn, such as $\ket{\uparrow, \rightarrow}$ and pairs related by a counterclockwise one, such as $\ket{\uparrow, \leftarrow}$. We do not do this in order to preserve the underlying symmetry of the non-interacting search. Specifically, the Grover coin treats all four directions equally and the lattice, as well as $S_{\mathrm{ff}}$, looks the same after a rotation by any multiple of $90^\circ$ or a reflection across a horizontal, vertical or diagonal axis, provided the coin directions are relabeled accordingly. Similarly, $\ket{\Psi_0}$ is unchanged by these transformations and so is $\mathcal{C}'$ if they are taken about the marked vertex $\mathbf{m}$, which is always possible with periodic boundary conditions. Since any perpendicular pair can be mapped onto any other by one of these transformations, an interaction that respects them must assign a single phase $\lambda_{\perp}$ to all eight. Rotations alone would not be enough, since they preserve the chirality of a pair: rotating $\ket{\uparrow, \rightarrow}$, for example, always produces another clockwise pair and only a reflection connects the two subclasses.

Exponentiating \cref{eq:ham_three_class} then results in
\begin{equation}
    \mathcal{O}_{ab} = I_c^{\otimes2}+z_{\parallel} P_{\parallel}^{(ab)} + z_{\rightleftarrows} P_{\rightleftarrows}^{(ab)} + z_{\perp} P_{\perp}^{(ab)},
\end{equation}
where $z_g=e^{i\lambda_g}-1$ for $g\in\left\{\parallel,\rightleftarrows,\perp\right\}$. The total generalized pairwise interaction operator for local phase interactions is therefore
\begin{equation}
    \mathcal{V}(\mathbf{\Lambda})=\prod_{a<b} \left[I+z_{\parallel} \Pi_{\parallel}^{(ab)} + z_{\rightleftarrows} \Pi_{\rightleftarrows}^{(ab)} + z_{\perp} \Pi_{\perp}^{(ab)}\right],
\end{equation}
where we have defined $\Pi_g^{(ab)}=P_g^{(ab)}\otimes\Pi_{ab}$ for $g\in\left\{\parallel,\rightleftarrows,\perp\right\}$ and $\mathbf{\Lambda} = \left(\lambda_{\parallel}, \lambda_{\rightleftarrows}, \lambda_{\perp}\right)$. Since each factor is diagonal in the joint direction and position basis of all $P$ walkers, the $\mathcal{O}_{ab}$ commute and the ordering ambiguity discussed after \cref{eq:general_P_interaction} disappears for this family of interactions.

\subsection{Restriction to a coin-flipping interaction}
\label{subsec:hpp_inter}

\begin{figure}[tb]
 \centering
 \includegraphics[width=\columnwidth]{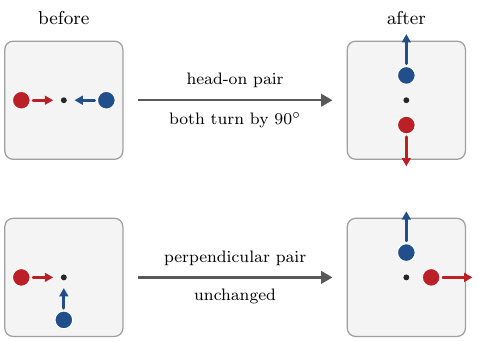}
 \caption{HPP collision rule for two particles meeting at a lattice site. Top: a head-on pair is scattered into the transverse axis, rotating both velocities by $90^\circ$. Bottom: a pair approaching along perpendicular directions retains its original velocities. Colors distinguish the particles, arrows indicate their velocities and the black dot marks the collision site.}
 \label{fig:hpp_collision_rule}
\end{figure}

The Hardy--Pomeau--de Pazzis (HPP) model~\cite{HarPoPaz1973} is one of the simplest lattice-gas automata describing a two-dimensional gas. In this model, each vertex of the two-dimensional grid carries four cells, one per cardinal direction, each holding at most one particle. Whenever two particles arriving from opposite directions meet at the same vertex, a collision occurs. If instead the particles arrive from perpendicular directions, no collision takes place and each particle is simply shifted or, in the language of HPP, streamed along its direction of motion. Overall, the collision rule leaves every configuration unchanged except for the head-on encounter, as depicted in \cref{fig:hpp_collision_rule}. This classical interaction model is particularly interesting and has already been studied by Costa, Melo and Portugal~\cite{CosMePor2019} for multi-particle quantum walk systems, because it is local and reversible and it conserves particle number and momentum at every vertex. It can thus be readily promoted to an analogous unitary operator for the coined quantum walk. Furthermore, unlike the simpler phase interaction formalized in the previous section, it flips the coin states of the walkers, giving us a broader picture of how different types of local coupling affect our model.

In the quantum walk setting, the four cells and their well-defined classical directions play the role of the coin basis~$\left\{\ket{\uparrow},\ket{\rightarrow},\ket{\leftarrow},\ket{\downarrow}\right\}$, while streaming corresponds, as already mentioned, to the shift operator. The collision event, in turn, becomes the local interaction operator $\mathcal{O}_{12}$ between the walkers. Unlike the phase interaction, however, we restrict the HPP interaction to $P=2$, since a head-on collision is intrinsically a two-body event and there is no canonical way to define the joint collision of three or more coincident walkers.

Conveniently, the head-on coin states on which the collision acts have already appeared in the previous section: they are exactly the antiparallel states, which span the head-on subspace $\mathcal{S}_{\mathrm{HO}}$ with projector
\begin{equation}
    P_{\mathrm{HO}} = \sum_{\mathbf{d}_1, \mathbf{d}_2} \delta_{\bar{\mathbf{d}}_1, \mathbf{d}_2} \ketbra{\mathbf{d}_1, \mathbf{d}_2}{\mathbf{d}_1, \mathbf{d}_2} = P_{\rightleftarrows}^{(12)}.
    \label{eq:head_on_proj}
\end{equation}
The intended action of the head-on rule is then to rotate both coins by $90^\circ$ in the same sense. In the coin encoding $\ket{c_x, c_y}$, a clockwise rotation of a single coin is the relabeling $\ket{c_x, c_y} \mapsto \ket{c_y, \bar{c}_x}$, which is implemented by a $\mathrm{SWAP}$ of the two coin qubits followed by an $\mathrm{X}$ gate on the second one. On $\mathcal{S}_{\mathrm{HO}}$, the collision is therefore
\begin{equation}
    C_{\mathrm{HO}} = (\mathrm{X}_{c_{y_1}} \mathrm{SWAP}_{c_{x_1}c_{y_1}}) \otimes (\mathrm{X}_{c_{y_2}} \mathrm{SWAP}_{c_{x_2}c_{y_2}}),
    \label{eq:coll_ho}
\end{equation}
which cycles through the four head-on states,
\begin{equation}
    \ket{\uparrow\downarrow} \mapsto \ket{\rightarrow\leftarrow} \mapsto \ket{\downarrow\uparrow} \mapsto \ket{\leftarrow\rightarrow} \mapsto \ket{\uparrow\downarrow}.
    \label{eq:ho_cycle}
\end{equation}
Since every other coin configuration is left untouched, the full collision operator and the corresponding interaction operator, i.e., \cref{eq:pair_interaction} with $\mathcal{O}_{12} = C_{\mathrm{coll}}$, read
\begin{align}
    C_{\mathrm{coll}} &= I_c^{\otimes 2} + \left(C_{\mathrm{HO}} - I_c^{\otimes 2}\right) P_{\mathrm{HO}}, \label{eq:coll_hpp_coin} \\
    V_{\mathrm{HPP}} &= I + \left(C_{\mathrm{coll}} - I_c^{\otimes 2}\right) \otimes \Pi_{12}. \label{eq:V_HPP}
\end{align}
Although $C_{\mathrm{HO}}$ is a product of single-walker rotations, conditioning it on the head-on subspace makes $C_{\mathrm{coll}}$ a two-particle unitary, which can entangle the walkers.

To then study the HPP interaction in more detail, we vary its strength. We thus promote the collision to a one-parameter family $C_{\mathrm{coll}}(\alpha)=e^{i\alpha H_{\mathrm{coll}}}$, which is unitary and reduces to the identity (the non-interacting case) at $\alpha=0$ and to the full collision at $\alpha=1$. Since the collision enters only through the interaction operator, it suffices to replace $C_{\mathrm{coll}} \to C_{\mathrm{coll}}(\alpha)$ in \cref{eq:V_HPP}. Taking into account then that the collision acts trivially outside the head-on subspace, the generator must vanish there, so $H_{\mathrm{coll}} = h_{\mathrm{HO}}P_{\mathrm{HO}}$, with $e^{ih_{\mathrm{HO}}}=C_{\mathrm{HO}}$ on $\mathcal{S}_{\mathrm{HO}}$. This leaves us with
\begin{equation}
    C_{\mathrm{coll}}(\alpha)=I_c^{\otimes 2}+\left(C_{\mathrm{HO}}^{\alpha}-I_c^{\otimes 2}\right)P_{\mathrm{HO}}.
    \label{eq:coll_alpha}
\end{equation}
By \cref{eq:ho_cycle}, $C_{\mathrm{HO}}$ acts as a cyclic shift on the head-on states $\ket{g_0} = \ket{\uparrow\downarrow}$, $\ket{g_1} = \ket{\rightarrow\leftarrow}$, $\ket{g_2} = \ket{\downarrow\uparrow}$, $\ket{g_3} = \ket{\leftarrow\rightarrow}$, i.e., $C_{\mathrm{HO}} \ket{g_n} = \ket{g_{n+1 \bmod 4}}$. It is therefore diagonalized by the Fourier modes
\begin{equation}
    \ket{v_k} = \frac{1}{2} \sum_{n=0}^{3} i^{-kn} \ket{g_n}, \qquad C_{\mathrm{HO}} \ket{v_k} = i^{k} \ket{v_k},
\end{equation}
with $k = 0, 1, 2, 3$. Taking the eigenphases $k\pi/2 \in [0, 2\pi)$ to define the fractional power, we obtain on $\mathcal{S}_{\mathrm{HO}}$
\begin{subequations}
\label{eq:frac_power}
\begin{align}
    C_{\mathrm{HO}}^{\alpha} &= \sum_{k=0}^{3} e^{i\alpha k\pi/2} \ketbra{v_k}{v_k} = \sum_{n=0}^{3} c_n(\alpha)\, C_{\mathrm{HO}}^{n},
    \\
    c_n(\alpha) &= \frac{1}{4} \sum_{k=0}^{3} e^{i(\alpha - n)k\pi/2}.
\end{align}
\end{subequations}
For integer $\alpha = m \in \{0,1,2,3\}$, $c_n(m) = \delta_{nm}$, so $C_{\mathrm{HO}}^{\alpha}$ passes through $I_c^{\otimes 2}$, $C_{\mathrm{HO}}$, $C_{\mathrm{HO}}^2$ and $C_{\mathrm{HO}}^3$. Moreover, $c_n(\alpha+4) = c_n(\alpha)$, so $C_{\mathrm{HO}}^{\alpha}$ has period 4 in $\alpha$.

At this point, we can see that the relevant range of the interaction strength is $\alpha \in [0, 4)$. This range can be narrowed further with a symmetry argument. Both walkers start in the same state and every operator in the walk, i.e., the coin $G$, the shift $S_{\mathrm{ff}}$, the marked-subspace projector $\mathcal{P}_m$, the same-position projector $\Pi_{12}$ and $C_{\mathrm{coll}}(\alpha)$, commutes with the exchange of the two walkers, so the joint state remains symmetric under particle exchange at all times. At coincident positions, the exchange acts on the coins alone and only the exchange-symmetric head-on coin states are ever populated. These are spanned by the head-on pairs along each axis,
\begin{equation}
\begin{aligned}
    \ket{\phi_y} &= \frac{1}{\sqrt{2}}\left(\ket{\uparrow\downarrow} + \ket{\downarrow\uparrow}\right), \\
    \ket{\phi_x} &= \frac{1}{\sqrt{2}}\left(\ket{\rightarrow\leftarrow} + \ket{\leftarrow\rightarrow}\right),
\end{aligned}
\label{eq:phi_states}
\end{equation}
which $C_{\mathrm{HO}}$ simply exchanges, $C_{\mathrm{HO}} \ket{\phi_{x,y}} = \ket{\phi_{y,x}}$. This is exactly the classical HPP rule and on this subspace $C_{\mathrm{HO}}^2$ is the identity, recovering the involuntory property of the classical collision. Consequently, the antisymmetric modes $\ket{v_1}$ and $\ket{v_3}$ drop out, leaving $\ket{v_0} = (\ket{\phi_y} + \ket{\phi_x})/\sqrt{2}$ with eigenvalue $+1$ and $\ket{v_2} = (\ket{\phi_y} - \ket{\phi_x})/\sqrt{2}$ with eigenvalue $-1$. On exchange-symmetric states, \cref{eq:V_HPP,eq:coll_alpha,eq:frac_power} therefore reduce to
\begin{equation}
    V_{\mathrm{HPP}}(\alpha) = I + \left(e^{i\pi\alpha} - 1\right) \ketbra{v_2}{v_2} \otimes \Pi_{12}.
    \label{eq:V_HPP_symmetric}
\end{equation}

Three consequences follow. First, the dynamics have period 2 in $\alpha$: $\alpha = 1$ is the full collision and $\alpha = 2$ reproduces the non-interacting walk exactly, so $\alpha \in [0, 2)$ covers every distinct interaction strength. Second, in this subspace the HPP interaction is itself a phase interaction, of the same form as the pairwise operator $V_{ab}$ of \cref{eq:pair_interaction}. The only difference from \cref{subsec:phase_inter} is that the phase acts on the entangled coin state $\ket{v_2}$, a superposition of head-on pairs along the two axes, rather than on a subspace diagonal in the joint direction basis. The two interaction models thus differ only in which coincident coin states receive the phase. Third, $C_{\mathrm{HO}}$ commutes with the rotations of both coins, while the reflections reverse the sense of rotation and map $C_{\mathrm{HO}}$ to $C_{\mathrm{HO}}^{-1}$. The two coincide on the exchange-symmetric subspace, where $\ketbra{v_2}{v_2}$ is invariant under all of $D_4$, the symmetry group of the square lattice discussed in \cref{subsec:phase_inter}. The clockwise convention in \cref{eq:coll_ho} is therefore immaterial for our search and the HPP interaction shares the $D_4$ symmetry of the phase interaction.

Finally, since a collision is always followed by streaming in the HPP model, we apply the interaction between the perturbed search coin $\mathcal{C}'$ of \cref{eq:perturbed_multi_coin} and the shift,
\begin{equation}
    \mathcal{U}_{\mathrm{HPP}}(\alpha) = S_{\mathrm{ff}}^{\otimes 2}\, V_{\mathrm{HPP}}(\alpha)\, \mathcal{C}',
    \label{eq:hpp_search_evol}
\end{equation}
in contrast with the phase interaction, which we apply after each non-interacting step of the walk. 

\section{Halo-exchange algorithm for simulating quantum walks}
\label{sec:halo_exchange}

The main obstacle we faced while attempting to simulate larger lattices and more particles was the size of the system state. The joint state of the $P$ walkers contains $(4N)^P$ complex amplitudes and offers no structure that would allow a compact representation. This is due to the particles initially being in a uniform superposition, so all of its entries are nonzero, which makes sparse storage no more memory-efficient. Additionally, the walkers become entangled during the search, so the state cannot be stored as a product of single-walker states. At \texttt{complex128} precision, three walkers on a $20\times20$ lattice already require approximately $65.5$~GB for a single copy of the state. In serial runs on a machine with 32~GB of RAM, the largest three-walker lattice that fit in memory, together with the working arrays of the time evolution, was $L=10$.

Several approaches to parallelizing quantum walk simulations already exist. For example, parallel Fortran libraries have been explored for continuous-time quantum walk simulations relying on sparse data structures~\cite{MatWang2021a,MatWang2021b} which, as argued above, offer no advantage here. GPU-accelerated algorithms have also been developed for continuous-time walks~\cite{PicBeSi2017}, as well as a heterogeneous computing approach covering both simulations of continuous- and discrete-time walks~\cite{MotBeSan2023}. In our case, the walk evolves on a square lattice, which makes a domain decomposition of the lattice a natural way to distribute the state. Since neither the phase nor the HPP interaction changes the runtime of the search (see Sec.~\ref{sec:interaction_results}), we use the distributed implementation only for the non-interacting search. In that walk, the Grover coin acts only on the coin axes and the marked coin $\mathcal{C}'$ is diagonal in position, so the flip-flop shift $S_{\mathrm{ff}}$ is the only operator that couples different positions, but only nearest-neighbor vertices. Updating a site therefore requires data only from its immediate neighbors, which is the same setup as for halo exchange~\cite{Quinn2003,MatSanMas2004} -- a standard method for nearest-neighbor stencil computations. Halo exchange involves splitting the lattice into blocks, one per process and, before each update, every process receives a one-site-wide strip of boundary data, the halo, from each neighboring block. Communication thus scales with the perimeter of a block, while computation scales with its area.

We apply this decomposition to walker~1's lattice only and distribute it over the MPI ranks of the DelftBlue cluster~\cite{DHPC2024}. Splitting this single axis already divides the state among $R$ ranks into chunks of approximately $(4N)^P/R$ amplitudes, while the shifts of walkers $2,\ldots,P$ remain entirely local. The coin and oracle steps require no communication and only walker~1's shift involves the halo exchange. The following subsections describe the domain decomposition and the boundary exchange. With this implementation, we extended the three-walker simulations of Sec.~\ref{sec:search_results} from $L=10$ to $L=25$.

\subsection{Domain decomposition}

We arrange the $R$ available MPI ranks on a two-dimensional Cartesian process grid of size $p_x\times p_y=R$. Following Fig.~\ref{fig:halo_exchange}(a), we take $x$ as the row index, increasing downward and $y$ as the column index, increasing to the right. The coin states $\ket{\rightarrow}$ and $\ket{\leftarrow}$ thus step along $\pm\hat x$, while $\ket{\uparrow}$ and $\ket{\downarrow}$ step along $\pm\hat y$. The rank with grid coordinates $(r_x,r_y)$ owns the block of walker~1's lattice
\begin{equation}
 \mathcal B_{r_x,r_y}=\{\mathbf{x}:\,\mathbf{x}_0\leq \mathbf{x}<\mathbf{x}_0+B_\mathbf{x}\},
 \label{eq:halo_block}
\end{equation}
where $\mathbf{x}=(x,y)$, $\mathbf{x}_0=(x_0,y_0)$ is the origin of the block and $B_\mathbf{x}=(B_x,B_y)$ is its width and height. The block height $B_x$ equals $\lfloor L/p_x\rfloor$ or $\lfloor L/p_x\rfloor+1$, with the first $L\bmod p_x$ process rows taking the extra siteand likewise for $B_y$. The local state tensor then has shape
\begin{equation}
 (4,\,B_xB_y,\,4,\,N,\,\ldots,\,4,\,N),
 \label{eq:halo_local_shape}
\end{equation}
where $B_xB_y$ replaces $N$ on walker~1's position axis. Each rank retains the full position ranges of walkers $2,\ldots,P$, conditioned on its own block of walker~1, so that the ranks partition the state without duplicating any amplitude.

With this decomposition, every operation other than walker~1's shift is local. To see this, we write the state as a tensor in index notation,
\begin{equation}
    \ket{\Psi} \equiv \Psi_{c_1p_1c_2p_2\ldots c_Pp_P},
\end{equation}
where $c_a$ and $p_a$ denote the coin and position indices of walker~$a$. At every vertex except $\mathbf{m}$, the Grover coin acts on each walker's coin axis independently. The full $P$-walker coin therefore never has to be constructed explicitly. Instead, $G$ is contracted with one coin axis at a time. For walker~1, for example,
\begin{equation}
    \Psi'_{k_1p_1c_2p_2\ldots c_Pp_P}=G_{k_1l_1}\Psi_{l_1p_1c_2p_2\ldots c_Pp_P},
\end{equation}
where repeated indices -- also known as ``dummy indices" -- are summed over. Since this contraction leaves every position index untouched, each rank applies it to its local tensor without any communication.

The full perturbed coin $\mathcal{C}'$ of Eq.~\eqref{eq:perturbed_multi_coin} is diagonal in position and acts as $-I_c^{\otimes P}$ on every configuration with at least one walker at $\mathbf m$. After applying the Grover contraction for each walker, each rank therefore overwrites every slice with $\mathbf x_a=\mathbf m$, for any walker $a$, with the negated pre-coin amplitudes. Because these slices are overwritten rather than multiplied by $-1$ once per walker, a configuration with $k>1$ walkers at $\mathbf m$ receives a single $\pi$ phase, as $\mathcal{C}'$ requires, rather than $(-1)^k$. For walkers $2,\ldots,P$, every rank holds its part of the marked slice and can index it globally. Conversely, for walker~1 only the rank whose block contains $\mathbf m$ holds the slice, which it locates through a precomputed map from global to local site indices.

\begin{figure}[tb]
    \centering
    \includegraphics[width=\columnwidth]{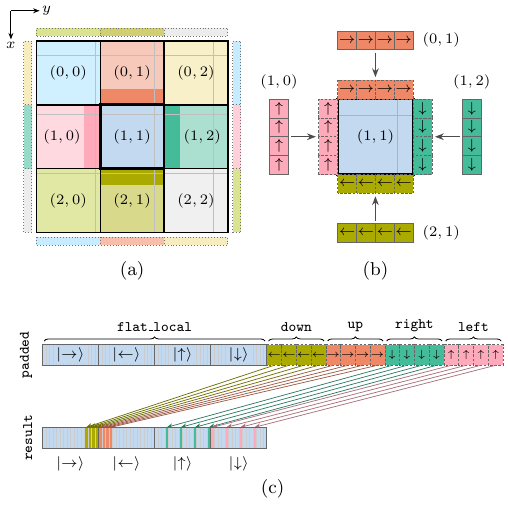}
    \caption{Halo exchange for the distributed $P$-walker search. (a) Walker~1's lattice distributed over a $3\times3$ process grid. The position axes of the remaining walkers are kept whole on every rank. The row index $x$ increases downward and the column index $y$ to the right, so $\ket{\rightarrow}$ and $\ket{\leftarrow}$ step along $\pm\hat x$ and $\ket{\uparrow}$ and $\ket{\downarrow}$ along $\pm\hat y$. Dotted bands mark the periodic boundary, colored by the rank whose data wrap to that side. (b) Rank $(1,1)$ receives one boundary strip from each cardinal neighbor, each carrying the single coin component that points into its blockand sends the corresponding outward-pointing strips. (c) The halo buffers are appended to the local block along walker~1's flattened coin-position axis and the precomputed permutation $\pi_{\mathrm{loc}}$ gathers the boundary entries of the shifted state from the halo strips and all other entries from the local block.}
    \label{fig:halo_exchange}
\end{figure}

\subsection{Boundary exchange and padded shift}

Regarding the flip-flop shift (Eq.~\eqref{eq:flipflop_shift}), since it is an involution, each amplitude of the shifted state is read from a single source,
\begin{equation}
 \Phi'_{\mathbf d,\mathbf x}=\Phi_{\bar{\mathbf{d}},\,\mathbf x+\mathbf d},
 \label{eq:halo_source}
\end{equation}
where we display only walker~1's indices. The amplitude with coin $\mathbf d$ at $\mathbf x$ thus comes from the neighboring site in direction $\mathbf d$, with the opposite coin. For interior sites of a block, this source is local. For a site on the edge of the block that faces direction $\mathbf d$, the source lies in the neighboring block. A new $\ket{\rightarrow}$ amplitude on the bottom row, for example, is read from the $\ket{\leftarrow}$ amplitude on the top row of the rank below. Each rank therefore exchanges one strip with each of its four cardinal neighbors, receiving the coin component pointing into its block and sending the one pointing out of it (Fig.~\ref{fig:halo_exchange}(b)). For the ranks on the edges of the process grid, these neighbors are periodic, which implements the boundary conditions. It is also important to note that each strip contains the full positions of walkers $2,\ldots,P$, since only walker~1's lattice is partitioned.

Once the exchange completes, the rank appends the four received strips to its local block along walker~1's flattened coin-position axis, forming the padded array of Fig.~\ref{fig:halo_exchange}(c). A local permutation $\pi_{\mathrm{loc}}$ then maps each output index either into the local block or into one of the halo strips. We build $\pi_{\mathrm{loc}}$ once at initialization by evaluating the global flip-flop permutation $\pi$ at every local site and redirecting the sources that fall outside the block into the corresponding strip. Walkers $2,\ldots,P$ use $\pi$ directly, so the complete shift $S_{\mathrm{ff}}^{\otimes P}$ reduces to a single gather from the padded array, indexed by $\pi_{\mathrm{loc}}$ on walker~1's axis and by $\pi$ on all others. Algorithm~\ref{alg:halo_step} summarizes one distributed step and the implementation is available in Ref.~\cite{MeOri2026}.

\begin{algorithm}[tb]
    \caption{One step of the distributed $P$-walker search on a single rank.}
    \label{alg:halo_step}
    \begin{algorithmic}[1]
    \Require local tensor $\phi$ of shape~\eqref{eq:halo_local_shape}; neighbor ranks; permutations $\pi_{\mathrm{loc}}$ and $\pi$
    \Ensure $\phi$ after one search step
    \State $\phi_0\gets\phi$
    \For{$a=1,\ldots,P$}
    \State apply $G$ to the coin state of walker $a$
    \EndFor
    \For{$a=1,\ldots,P$} \Comment{marked coin}
    \State set $\phi\gets-\phi_0$ on the local slice $\mathbf x_a=\mathbf m$
    \EndFor
    \State send the outgoing strips to the four neighbors: top-row $\ket{\leftarrow}$ up, bottom-row $\ket{\rightarrow}$ down, left-column $\ket{\downarrow}$ left, right-column $\ket{\uparrow}$ right
    \State receive the incoming strips $h_{\mathrm{down}},h_{\mathrm{up}},h_{\mathrm{right}},h_{\mathrm{left}}$
    \State $\tilde\phi\gets(\phi\,|\,h_{\mathrm{down}}\,|\,h_{\mathrm{up}}\,|\,h_{\mathrm{right}}\,|\,h_{\mathrm{left}})$ along walker~1's flattened axis
    \State $\phi\gets\tilde\phi[\pi_{\mathrm{loc}},\pi,\ldots,\pi]$ \Comment{$S_{\mathrm{ff}}^{\otimes P}$}
    \end{algorithmic}
\end{algorithm}

The single-walker partition is also the building block for more general simulations. Partitioning every walker's lattice, each on its own pair of process-grid dimensions, amounts to repeating the exchange and padded gather for each walker's shift. Any additional operator that is diagonal in position, such as the local interactions of Sec.~\ref{subsec:interactions}, remains local under either partition and can be inserted into the step without further communication.

\section{Non-interacting quantum walkers: numerical results}
\label{sec:search_results}

In this section, we present numerical results for the multi-particle quantum walk search algorithm introduced in~\Cref{sec:model}. We first validate the dynamics of the $P$-particle search for $P>1$ and then use the von Neumann entropy to quantify the entanglement between the particles. Before doing so, we describe how the success probabilities and entropies are extracted from the simulations and define the quantities studied in the following sections.

\subsection{Observables and simulation protocol}
\label{subsec:observables}

Every walk, whether single- or multi-particle, is run for a fixed number of steps $T_{\mathrm{win}}$. For a single particle, the optimal number of steps is known to scale as $\sqrt{N\log N}$~\cite{AmKemRi2004,Tul2008}, so it suffices to simulate up to $T_{\mathrm{win}}=k\sqrt{N\log N}$ for some constant integer $k>0$. Throughout this work, $\log$ denotes the natural logarithm unless a base is specified. For the multi-particle search, no analytical estimate of the runtime is available yet, so we use the same window. Grover's lower bound guarantees that no search algorithmand hence not our walk either, can find the marked vertex in fewer than $\Omega(\sqrt{N})$ steps. Since $\sqrt{N\log N}$ grows faster than $\sqrt{N}$, the window would therefore still contain the success peak even if the multi-particle algorithm were to reach the Grover bound. That the window is also long enough, i.e., that the multi-particle search is not slower than the single-particle one, we verify a posteriori: in all reported runs, the first maximum of the success probability lies inside the window. We define the peak time $T_{\mathrm{peak}}$ as the step at which the success probability $p_{\mathrm{succ}}(T)$ is maximal within its first period, i.e., before it first returns close to its initial value.

To describe the spatial distribution of the walkers, we consider the probability mass function of the $P$ particles over the lattice, whose vertices we label by their position $\mathbf{x}$. We call the probability that at least one particle occupies $\mathbf{x}$ the any-walker probability $p_{\mathrm{any}}(\mathbf{x})$. Because the search succeeds whenever at least one walker is found at the marked vertex (see~\Cref{sec:quantum-search-problem}), the success probability is $p_{\mathrm{succ}}=p_{\mathrm{any}}(\mathbf{m})$. Both quantities depend implicitly on the step $T$, i.e., $p_{\mathrm{any}}(\mathbf{x};T)$ and $p_{\mathrm{succ}}(T)=p_{\mathrm{any}}(\mathbf{m};T)$. This allows us to define the peak success probability $p_{\mathrm{peak}}=p_{\mathrm{succ}}(T_{\mathrm{peak}})$.

Finally, as already mentioned, both the joint perturbed coin $\mathcal{C}'$ and the interactions introduced in~\Cref{subsec:interactions} can entangle the walkers. We quantify this entanglement with the von Neumann entropy of a single walker. After $T$ steps, the full system is in the pure state
\begin{equation*}
    \rho_{12\ldots P}(T)=\ketbra{\Psi_T}{\Psi_T}=\mathcal{U}^T\ketbra{\Psi_0}{\Psi_0}\big(\mathcal{U}^{\dagger}\big)^T.
\end{equation*}
Tracing out the coin and position degrees of freedom of all walkers except walker 1 gives the reduced density matrix $\rho_1=\Tr_{2,\ldots,P}\ketbra{\Psi_T}{\Psi_T}$, whose entropy is
\begin{equation}
    S(\rho_1)=-\Tr(\rho_1\log_2\rho_1).
    \label{eq:entropy}
\end{equation}
It suffices to consider walker 1. The initial state $\ket{\Psi_0}$ is symmetric under the exchange of any two walkers and the evolution operator $\mathcal{U}$ commutes with these exchanges, so $\ket{\Psi_T}$ remains exchange symmetric at every step. All single-walker reduced density matrices are therefore identical and $S(\rho_1)$ measures the entanglement of any one walker with the remaining $P-1$. Since the joint state is pure, $S(\rho_1)=0$ if and only if walker 1 is not entangled with the othersand the entropy grows as $\rho_1$ becomes more mixed. It is maximal when $\rho_1$ is maximally mixed, i.e., when walker 1 is maximally entangled with the rest. On a grid of $N=L\times L$ vertices, the single-walker Hilbert space has dimension $4N$, with four coin states per vertex, so that $S(\rho_1)\leq\log_2(4N)$.

\subsection{Multi-particle search dynamics}

\begin{figure}[tb]
    \centering
    \includegraphics[scale=1]{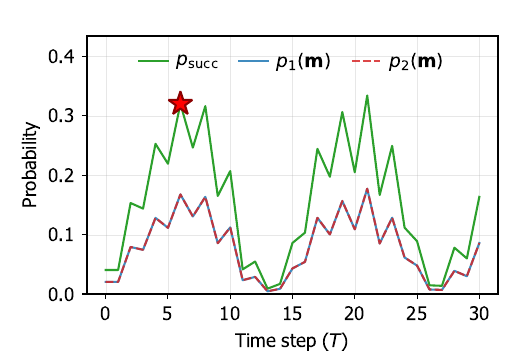}
    \caption{Two-walker probabilities at the marked vertex over the first 30 steps of the search, with $T_{\mathrm{peak}}$ represented by the star. The success probability $p_{\mathrm{succ}}=p_{\mathrm{any}}(\mathbf{m})$ is shown together with the marginal probabilities $p_a(\mathbf{m})$ of walker $a=1,2$ being at $\mathbf{m}$. The two marginals coincide because the state $\ket{\Psi_T}=\mathcal{U}^T\ket{\Psi_0}$ is exchange symmetric at every step $T$.}
    \label{fig:two_part_evol}
\end{figure}

\begin{figure}[tb]
    \centering
    \includegraphics[scale=1]{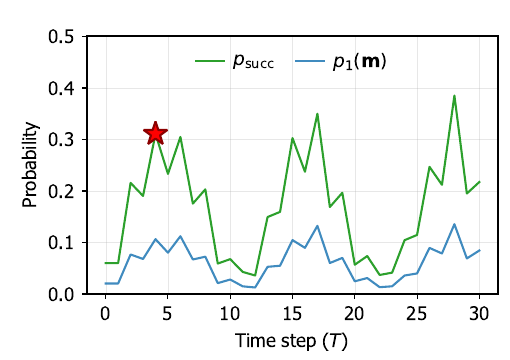}
    \caption{Same as~\Cref{fig:two_part_evol} for $P=3$. Only the marginal $p_1(\mathbf{m})$ is shown, since all three marginals coincide by exchange symmetry.}
    \label{fig:three_part_evol}
\end{figure}

We first run the non-interacting multi-particle search of~\Cref{sec:model} on a $7\times7$ lattice ($N=49$) with $P=1$, $2$ and $3$ walkers, placing the marked vertex at $\mathbf{m}=(2,2)$. With periodic boundary conditions, this choice is arbitrary, since all vertices are equivalent. \Cref{fig:two_part_evol} shows the success probability $p_{\mathrm{succ}}$ for $P=2$, together with the marginal probabilities $p_1(\mathbf{m})$ and $p_2(\mathbf{m})$ of each walker being at the marked vertex. The success probability oscillates quasi-periodically, with a period of approximately $14$ steps and, as expected, the two marginals are identical, since the state is exchange symmetric. Following~\Cref{subsec:observables}, we only consider the first period, where $T_{\mathrm{peak}}=6$ steps and $p_{\mathrm{peak}}\approx0.32$. For $P=3$, \Cref{fig:three_part_evol} shows the same behavior with a shorter period of about $11$ steps, $T_{\mathrm{peak}}=4$ and $p_{\mathrm{peak}}\approx0.31$.

\begin{figure*}[tb]
    \centering
    \includegraphics[scale=1]{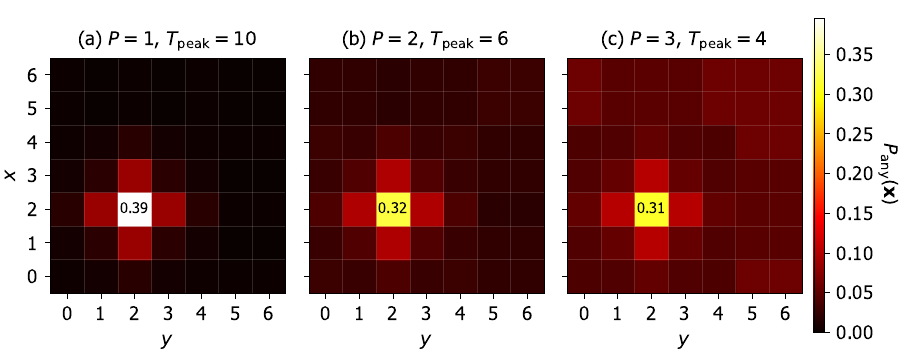}
    \caption{Any-walker probability $p_{\mathrm{any}}(\mathbf{x})$ on the $N=7\times7=49$ lattice for (a) $P=1$, (b) $P=2$ and (c) $P=3$, evaluated at the peak step $T_{\mathrm{peak}}$ within the first ``period'' of the evolution. The success probability $p_{\mathrm{succ}}=p_{\mathrm{any}}(\mathbf{m})$ is printed at the marked vertex $\mathbf{m}=(2,2)$. It is slightly lower for $P>1$ than for $P=1$, but still clearly exceeds the probability at all neighboring vertices, so the walkers remain localized at the target.}
    \label{fig:heatmaps_P123}
\end{figure*}

\Cref{fig:heatmaps_P123} shows the corresponding any-walker probabilities $p_{\mathrm{any}}(\mathbf{x})$ at $T_{\mathrm{peak}}$ for $P=1$, $2$ and $3$. In all three cases, the probability clearly localizes at $\mathbf{m}$. Adding walkers shortens the time to the first peak, from $T_{\mathrm{peak}}=10$ for $P=1$ to $6$ and $4$ for $P=2$ and $P=3$, respectively, but lowers the peak success probability from $0.39$ for $P=1$ to $0.32$ and $0.31$. As we show in~\Cref{sec:scaling}, this decrease is a small-grid effect: for $P=2$ and $L\gtrsim30$, the two-walker $p_{\mathrm{peak}}$ matches or slightly exceeds the single-walker value. For $P=3$, our simulations extend only up to $L=25$, so whether the same holds there remains to be confirmed. In addition, going from $P=1$ to $P>1$ increases the probability at the unmarked vertices, since $p_{\mathrm{any}}(\mathbf{x})$ there receives contributions from all $P$ walkers. Nevertheless, $p_{\mathrm{peak}}$ remains, at least,  $\sim50\%$ higher than the background vertices, so a constant number of rounds of amplitude amplification suffices to retrieve the marked vertex with high probability. Alternatively, Tulsi's algorithm~\cite{Tul2008} could be adapted to the multi-particle case, which we leave for future work.
 
We note also that $p_{\mathrm{succ}}$ in~\Cref{fig:two_part_evol,fig:three_part_evol} is jagged, alternating between even and odd steps, so that neighboring local maxima can have nearly equal heights, e.g., at $T=6$ and $T=8$ in the first period for $P=2$ and at $T=4$ and $T=6$ for $P=3$. This artificial near-degeneracy makes $T_{\mathrm{peak}}$ sensitive to small changes in the dynamicsand we return to it in~\Cref{sec:scaling}.

\subsection{Entanglement generation during the search}

\begin{figure}[tb]
    \centering
    \includegraphics[scale=1]{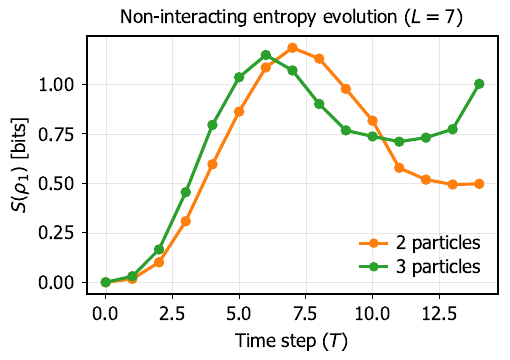}
    \caption{One-walker von Neumann entropy $S(\rho_1)$, in bits, of the non-interacting search on the $7\times7$ lattice for $P=2$ and $P=3$ over the first $14$ steps. The entropy vanishes for the initial product state and peaks one to two steps after the respective $T_{\mathrm{peak}}$, well below the bound $\log_2(4N)\approx7.6$ bits.}
    \label{fig:entropy_evol_P23}
\end{figure}

\Cref{fig:entropy_evol_P23} shows the one-walker entropy $S(\rho_1)$ for the non-interacting search with $P=2$ and $P=3$ on the same $7\times7$ lattice. Although no interaction has been added, the entropy grows from zero, since the joint coin $\mathcal{C}'$ applies a $\pi$ phase shift to the coins of all walkers whenever any of them is at $\mathbf{m}$ and thereby correlates their motion. The entropy rises together with the success probability and reaches its maximum, about $1.2$ bits, at $T=7$ for $P=2$ and at $T=6$ for $P=3$, one to two steps after the respective $T_{\mathrm{peak}}$. It then decreases as the probability moves away from the marked vertex. For $P=3$, it rises again towards the end of the window, as the success probability begins its second period (\Cref{fig:three_part_evol}). Entanglement is therefore generated mainly while the walkers localize at the targetand the entropy closely follows the success probability with a delay of one to two steps.

\begin{figure}[tb]
    \centering
    \includegraphics[scale=1]{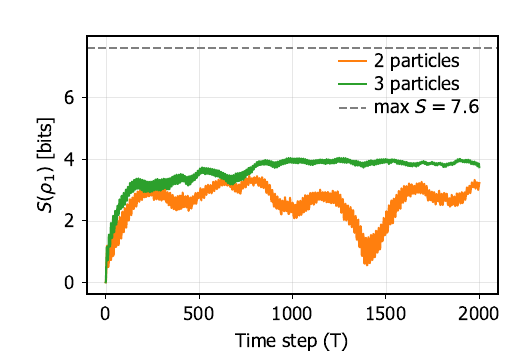}
    \caption{One-walker entropy $S(\rho_1)$, in bits, of the non-interacting search on the $7\times7$ lattice for $P=2$ and $P=3$ over $2000$ steps. The dashed line marks the upper bound $\log_2(4N)\approx7.6$ bits.}
    \label{fig:long_entropy}
\end{figure}

Additionally, the entropy does not return to zero after the first period. \Cref{fig:long_entropy} follows $S(\rho_1)$ over $2000$ steps. Within the first $200$ steps, it builds up to about $3$ bits for both $P=2$ and $P=3$. For $P=2$, it then fluctuates between roughly $2$ and $3.4$ bits, with a deep dip down to below $1$ bit at $T=1400$, while for $P=3$ it keeps growing slowly to about $4$ bits and its dips are shallower. In both cases, the entropy remains well below the bound $\log_2(4N)\approx7.6$ bits, so walker 1 never becomes maximally mixed. The search thus does not require a large amount of entanglement to localize the probability at the marked vertex. In~\Cref{sec:interaction_results}, we show that local interactions can generate more entanglement, but at the cost of a lower peak success probability.

\section{Effect of Local Interactions}
\label{sec:interaction_results}

The non-interacting case results establish that the joint search generates both localization and entanglement and this entanglement originates from the marked vertex operator. Adding a local interaction in the form of collision allows us to change the correlations between the walkers while keeping the spatial lattice and the marked coin fixed. We study the HPP and phase models of interactionsand analyze the respective success probability, peak time and one-walker entropy together. This makes it possible to separate the effect of interaction on how rapidly the search develops from its effect on how much of probability weight reaches the target.

\begin{figure*}[tb]
    \centering
    \begin{minipage}{0.49\textwidth}
    \centering\textbf{(a)}\quad HPP collision interaction ($P=2$, $L=7$)\\
    \includegraphics[scale=0.8]{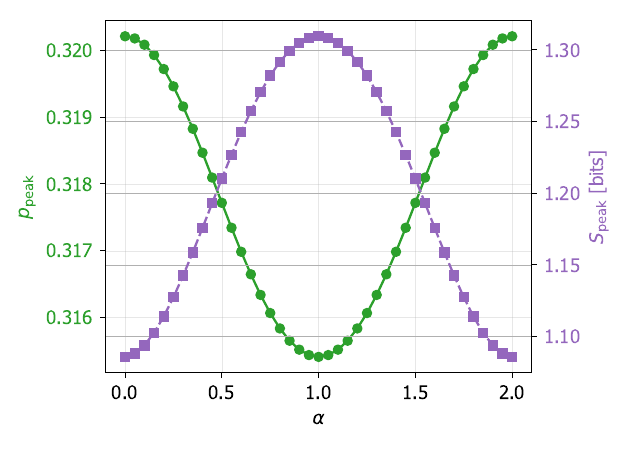}
    \end{minipage}\hfill
    \begin{minipage}{0.49\textwidth}
    \centering\textbf{(b)}\quad Uniform phase interaction ($P=2$, $L=7$)\\
    \includegraphics[scale=0.8]{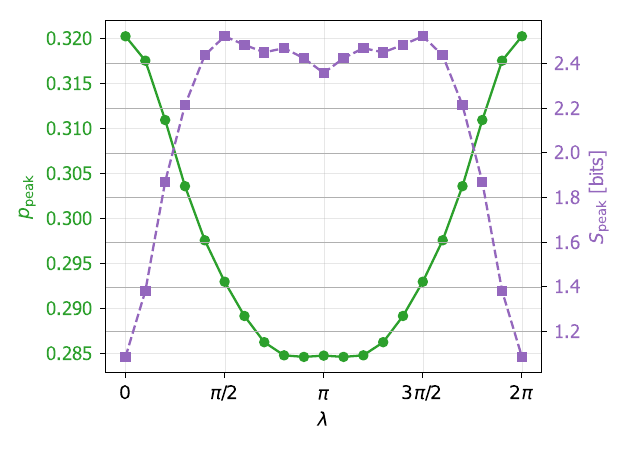}
    \end{minipage}
    \par\smallskip
    \begin{minipage}{0.49\textwidth}
    \centering\textbf{(c)}\quad HPP collision effect on runtime\\
    \includegraphics[scale=0.8]{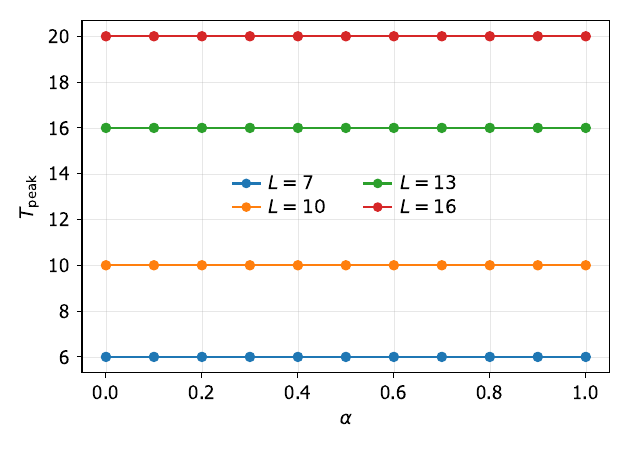}
    \end{minipage}\hfill
    \begin{minipage}{0.49\textwidth}
    \centering\textbf{(d)}\quad Phase interaction effect on runtime\\
    \includegraphics[scale=0.8]{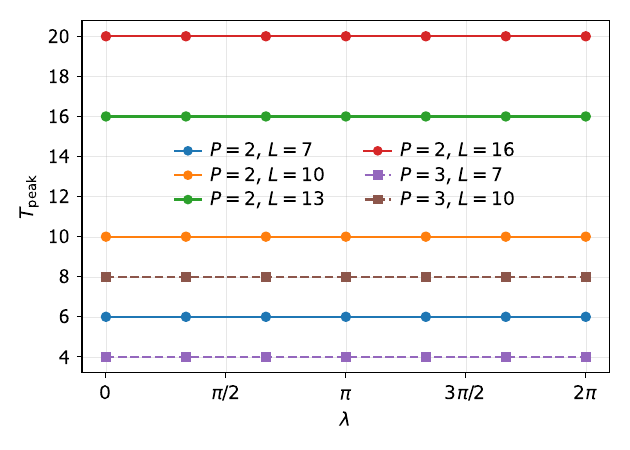}
    \end{minipage}
    \caption{Local interactions reduce peak success while generating additional entanglement. (a,b) Peak success probability $p_{\mathrm{peak}}$ and one-walker entropy $S_{\mathrm{peak}}$ at the first success peak for $P=2$ and $L=7$, versus (a) the HPP collision strength $\alpha$ and (b) the uniform phase $\lambda$. (c) $T_{\mathrm{peak}}$ versus $\alpha$ for $P=2$ and $L=7,10,13,16$. (d) $T_{\mathrm{peak}}$ versus $\lambda$ for $P=2$ at $L=7,10,13,16$ and $P=3$ at $L=7,10$, sampled at multiples of $\pi/3$.}
 \label{fig:interactions}
\end{figure*}

\subsection{HPP collisions in two-particle search}

\Cref{fig:interactions}(a,c) shows how the HPP collision affects the peak success probability $p_{\mathrm{peak}}$, the entropy at the peak, $S_{\mathrm{peak}}=S[\rho_1(T_{\mathrm{peak}})]$and the peak step $T_{\mathrm{peak}}$, our measure of the runtime, as functions of the collision strength $\alpha$. In panel (a), $p_{\mathrm{peak}}$ and $S_{\mathrm{peak}}$ are anticorrelated. As the collision is switched on and $\alpha$ increases from $0$ to $1$, the entropy grows by about $0.2$ bits, while the success probability decreases by only about $0.005$. The collision thus entangles the walkers further, but barely affects how much probability reaches the marked vertex. For $\alpha\in[1,2]$, both quantities return to their non-interacting values, as expected from the period of $2$ in $\alpha$.

Panel (c) shows that $T_{\mathrm{peak}}$ does not change with $\alpha$ for any of the simulated lattice sizes, $L=7$ to $16$. This holds even at $L=7$, where the two highest maxima in the first period differ by less than the change in $p_{\mathrm{peak}}$ caused by the collision. Within these sizes, the HPP collision therefore changes how much probability reaches $\mathbf{m}$, but not when it peaks.

\subsection{Uniform phase interactions}

We now consider the phase interaction of~\Cref{subsec:phase_inter} in the restricted form of a uniform phase, $\lambda_{\parallel}=\lambda_{\rightleftarrows}=\lambda_{\perp}=\lambda$, so that every coincident configuration acquires the same phase shift $e^{i\lambda}$. \Cref{fig:interactions}(b,d) shows its effect on the same observables. Both curves in panel (b) are symmetric about $\lambda=\pi$. This follows from complex conjugation, since $G$, $S_{\mathrm{ff}}$, $\mathcal{C}'$ and $\ket{\Psi_0}$ are real, so that the walk with phase $2\pi-\lambda$ is the complex conjugate of the walk with phase $\lambda$ and yields the same $p_{\mathrm{succ}}$ and $S(\rho_1)$.

The effect is considerably larger than for the HPP collision. As $\lambda$ increases from $0$ to $\pi/2$, the entropy at the peak more than doubles, from $1.09$ to $2.52$ bits, while $p_{\mathrm{peak}}$ decreases from $0.320$ to $0.293$. The success probability keeps decreasing and stays close to its minimum of $0.285$, about $11\%$ below the non-interacting value and seven times the decrease caused by the HPP collision, for $0.8\pi\lesssim\lambda\lesssim1.2\pi$. The entropy, in contrast, is maximal at $\lambda=\pi/2$ and $3\pi/2$ and has a shallow local minimum of $2.36$ bits at $\lambda=\pi$, so the anticorrelation seen for the HPP collision holds here only between $\lambda=0$ and $\pi/2$ and its mirror image. The larger effect is consistent with how the two interactions act on the exchange-symmetric states populated during the search. The HPP collision applies its phase only to the head-on state $\ket{v_2}$ of~\cref{eq:V_HPP_symmetric}, whereas the uniform phase acts on every coincident coin state.

The value $\lambda=\pi$ is special, since the interaction then multiplies every coincident configuration by $-1$, the same $\pi$ phase that $\mathcal{C}'$ imprints on the marked configurations. In the joint position space of~\Cref{fig:model}, the diagonal is thus phase-marked in addition to the row and column through $\mathbf{m}$, although without the replacement of the Grover coin performed by $\mathcal{C}'$. A possible explanation for the low $p_{\mathrm{peak}}$ around $\lambda=\pi$ is that this second set of phase-marked configurations draws amplitude away from the configurations with a walker at $\mathbf{m}$. Whether the same mechanism also causes the local minimum of the entropy remains to be tested.

Nevertheless, panel (d) shows that $T_{\mathrm{peak}}$ again does not depend on $\lambda$, for $P=2$ at $L=7$ to $16$ and for $P=3$ at $L=7$ and $10$. At $L=7$ and $\lambda=\pi$, the two highest maxima in the first period for $P=2$ differ by only $0.005$, much less than the $0.035$ decrease of $p_{\mathrm{peak}}$, so the unchanged peak step is not an artifact of the integer step resolution. Within the simulated sizes, neither interaction changes when the success probability peaks, only how high it gets. This is why we consider only the non-interacting search in~\Cref{sec:scaling}.

\section{Scaling with the database size}
\label{sec:scaling}

\begin{figure*}[tb]
    \centering
    \includegraphics[scale=1]{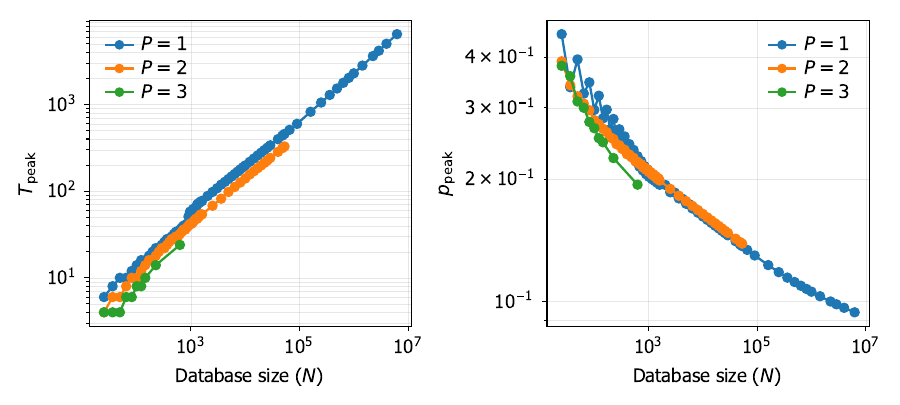}
    \caption{Search performance versus $N=L^2$ on logarithmic axes, with the peak step defined as the maximizing step $T_{\mathrm{peak}}$. Left: steps to the first success peak for $P=1,2,3$. Right: corresponding peak success probabilities. The simulated ranges are $5\leq L\leq2500$ for $P=1$, $5\leq L\leq230$ for $P=2$ and $5\leq L\leq25$ for $P=3$.}
    \label{fig:scaling_data}
\end{figure*}

In this section, we study how the runtime and the success probability scale with the database size $N$. \Cref{fig:scaling_data} shows $T_{\mathrm{peak}}$ and $p_{\mathrm{peak}}$ on logarithmic axes for $P=1$, $2$ and $3$, up to the largest lattices compatible with the halo exchange implementation of~\Cref{sec:halo_exchange} and the computing resources available on DelftBlue~\cite{DHPC2024}. For both $P=2$ and $P=3$, $T_{\mathrm{peak}}$ lies below the single-walker curve, indicating a speedup, although for $P=3$ this rests on a narrow range of small lattices. The peak success probability for $P=3$ stays below the single-walker value over the simulated range. For $P=2$, it is also lower at small sizes, but from $N\approx900$ onwards it matches or slightly exceeds the single-walker value. Whether the same happens for $P=3$ at larger sizes remains open.

\begin{figure}[tb]
    \centering
    \includegraphics[scale=1]{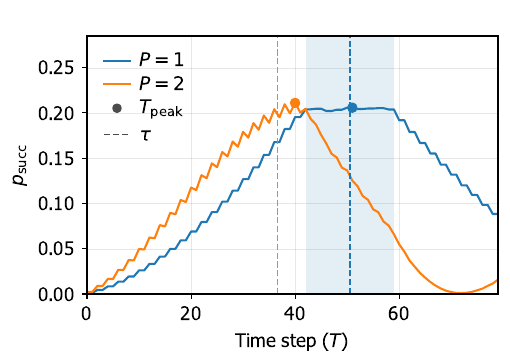}
    \caption{Success probability $p_{\mathrm{succ}}$ as a function of the time step $T$ for single- and two-walker search on a lattice of $N=900$ vertices, shown over one ``period'' of the search. Dots mark the maximizing step $T_{\mathrm{peak}}$ and dashed lines the fitted peak time $\tau$ of~\cref{eq:sin2_fit}. For $P=1$, the success probability stays within $3\%$ of its maximum over the shaded region, so $T_{\mathrm{peak}}$ is nearly degenerate, while $\tau$ lies in the middle of the plateau. For $P=2$, $p_{\mathrm{succ}}$ rises and falls without a plateau, so $T_{\mathrm{peak}}$ is better defined.}
    \label{fig:plateau_N900}
\end{figure}

The same sizes deserve a closer look for the runtime, since $T_{\mathrm{peak}}$ for $P=1$ jumps there, while for $P=2$ it keeps growing smoothly. \Cref{fig:plateau_N900} shows the first period of $p_{\mathrm{succ}}$ at $N=900$. For $P=1$, the success probability forms a long plateau near its maximum, so the step that maximizes it is nearly degenerate. Which step wins is decided by small kinks in $p_{\mathrm{succ}}$ that appear when the walk wraps around the periodic lattice, at $T\simeq\sqrt{2}\,d$, where $d=L,\sqrt{2}L,2L,\ldots$ are the distances from $\mathbf{m}$ to its periodic images. Below $N\approx900$, the maximum lies at the kink near $T\simeq\sqrt{2}L$, at the start of the plateauand above it at the kink near $T\simeq2L$, at its end. For $P=2$, $p_{\mathrm{succ}}$ rises and falls without a plateau, so $T_{\mathrm{peak}}$ stays near the first kink for all sizes. The ratio of the single- and two-walker peak steps therefore jumps around $N\approx900$ for a reason unrelated to the asymptotic scaling, which makes $T_{\mathrm{peak}}$ unsuitable for comparing the runtimes.

We therefore extract the runtime from the whole first lobe of $p_{\mathrm{succ}}(T)$ rather than from its maximum, by fitting
\begin{equation}
    p_{\mathrm{succ}}(T)\simeq A\sin^2\!\left(\frac{\pi T}{2\tau}\right)
    \label{eq:sin2_fit}
\end{equation}
over $0\leq T\leq 1.5\,\tau$ and iterating until the fit window and $\tau$ are consistent. The fitted $\tau$ is a real number, namely the step at which the smooth part of $p_{\mathrm{succ}}$ is maximaland $A$ is the corresponding height. At $N=900$, the fitted $\tau$ for $P=1$ lies in the middle of the plateau, as marked in~\Cref{fig:plateau_N900}. It is important to note, however, that since the walk operator is applied an integer number of times, $\tau$ serves only to characterize the scaling, while $T_{\mathrm{peak}}$ and $p_{\mathrm{peak}}$ remain the operational figures of merit at each simulated size.

 The form~\eqref{eq:sin2_fit} holds whenever the search probability behavior is dominated by the superposition of two eigenvectors $\ket{\phi_1}$ and $\ket{\phi_2}$ of $\mathcal{U}'$ with eigenphases $\phi_1$ and $\phi_2$. Let $Q_{\mathbf{m}}=I_c^{\otimes P}\otimes\mathcal{P}_{\mathbf m}$ project onto the configurations with at least one walker at $\mathbf m$, so that $p_{\mathrm{succ}}(T)=\lVert Q_{\mathbf{m}}\,\mathcal{U}'^{T}\ket{\Psi_0}\rVert^2$. If the initial state $\ket{\Psi_0}=c_1\ket{\phi_1}+c_2\ket{\phi_2}+\dots$ with $\dots$ representing all remaining eigenvectors that do not contribute significantly to search, the small initial value $p_{\mathrm{succ}}(0)=1-(1-1/N)^P$ forces the marked components to nearly cancel, $c_1Q_{\mathbf{m}}\ket{\phi_1}\simeq-c_2Q_{\mathbf{m}}\ket{\phi_2}\equiv\ket{a}$. Then
 \begin{align}
    p_{\mathrm{succ}}(T)&\simeq\lVert a\rVert^2\,\big|e^{i\phi_1T}-e^{i\phi_2T}\big|^2 \nonumber \\
    &=4\lVert a\rVert^2\sin^2\!\left(\frac{(\phi_1-\phi_2)\,T}{2}\right),
 \end{align}
 which is~\eqref{eq:sin2_fit} with $\tau=\pi/|\phi_1-\phi_2|$. The distribution of these eigenvectors in the spatial grid produces the short-scale structure described above. For a single walker, the dominance of such a pair underlies the analysis of Ref.~\cite{AmKemRi2004}. For $P\geq2$ we treat it as a model and test it against the simulations. Alternatively, one could also use the Fourier transform of the long-time probability evolution for extracting $\tau$~\cite{OriJinGre2024}. The deviation of the simulated $p_{\mathrm{succ}}$ from the fit decreases with $L$ for every $P$and for $P=2$ the fit also correctly predicts that $p_{\mathrm{succ}}$ returns to zero at $T=2\tau$, outside the fitted range.

\begin{figure}[tb]
    \centering
    \includegraphics[scale=1]{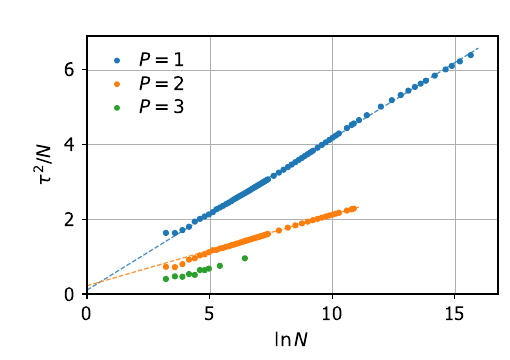}
    \caption{Squared fitted peak time divided by the database size, $\tau^2/N$, versus $\log N$ for $P=1,2,3$. Dashed lines are linear fits over $L\geq50$ for $P=1$ and $P=2$. A straight line corresponds to $\tau\propto\sqrt{N\log N}$ and the smaller slope for $P=2$ reflects a constant-factor reduction of the runtime.}
    \label{fig:scaling_asymptotic}
\end{figure}

We now compare $\tau$ with the single-walker scaling $\sqrt{N\log N}$~\cite{AmKemRi2004,Tul2008}. Rather than fitting reference curves on logarithmic axes, we plot $\tau^2/N$ against $\log N$ in~\Cref{fig:scaling_asymptotic}, where $\tau\propto\sqrt{N\log N}$ appears as a straight line and the Grover scaling $\tau\propto\sqrt{N}$ as a horizontal one. As expected, the $P=1$ points lie on a straight lineand so do the $P=2$ points, only with a smaller slope. The two-walker search therefore does not remove the $\log N$ factor to reach the Grover scaling. The $P=3$ points also rise with $\log N$, but they cover only small lattices, where finite-size effects are still strong.

\begin{figure}[tb]
    \centering
    \includegraphics[scale=1]{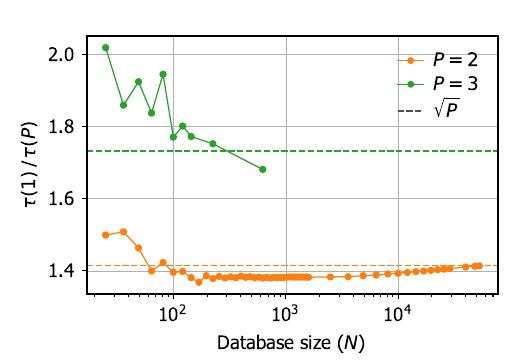}
    \caption{Ratio of single- to multi-walker fitted peak times, $\tau(1)/\tau(P)$, versus the database size $N$ for $P=2$ and $P=3$. Dashed lines mark $\sqrt{P}$. For $P=2$, the ratio approaches $\sqrt{2}$ slowly from below, while for $P=3$ it has not yet converged over the simulated range.}
    \label{fig:scaling_speedup}
\end{figure}

Although the scaling with $N$ is unchanged, $T_{\mathrm{peak}}$ in~\Cref{fig:scaling_data} is smaller for $P>1$, which suggests a constant-factor speedup. \Cref{fig:scaling_speedup} shows the ratio $\tau(1)/\tau(P)$, which for $P=2$ slowly approaches $\sqrt{2}$ as $N$ grows, consistent with the roughly halved slope in~\Cref{fig:scaling_asymptotic}. For $P=3$, the simulated lattices are too small to determine the limit. Our data thus suggest $\tau(P)\simeq\tau(1)/\sqrt{P}$, which we have established for $P=2$.

A constant factor does not change the complexity class, but it matters in practice. On platforms where a quantum walk is easier to implement than a gate-based Grover search, the number of walk steps sets the runtime. If the $\sqrt{P}$ trend extends to $P=4$, four walkers would halve the number of steps for large databases. Whether this also halves the physical runtime depends on how the duration of a single step and the required resources grow with $P$, which we do not discuss here as it might depend on the particular realization of a quantum simulator.

\section{Conclusions}
\label{sec:discussion}

In the present paper we studied the efficiency of a multi-particle quantum walk search algorithm on a two-dimensional grid. This is the only dimensionality in which the single-particle quantum walk search offers an advantage over its classical counterpart and, at the same time, does not reach Grover's search asymptotic of $O(\sqrt{N})$. Our study thus provides a step towards answering a fundamental question: is adding extra particles to the quantum walk search on the grid enough to achieve Grover's asymptotic speedup? We found the answer to be mixed: while the geometry of the grid preserves the $O(\sqrt{N\log N})$ scaling of the runtime, the extra degrees of freedom introduced by the additional particles allow us to attain a constant-factor improvement over the single-particle counterpart.

In the multi-particle setting, we also studied the effect of two local interactions, namely the phase and HPP interactions. We found that the search evolution operator entangles the walkers even in the absence of interactions. Switching on either interaction created additional inter-particle entanglement, but never increased the probability of finding the marked vertex. The interactions did, however, leave the overall structure of the localization at the marked vertex over time intact, so the runtime remained unchanged. We therefore did not need to study how the runtime scales with the database size in the presence of these interactions, as it would simply coincide with that of the non-interacting search.

Although adding particles does not change the inherent scaling of the 2D quantum walk search, it has practical value, as it reduces the runtime by a constant factor of $\sqrt{P}$. Multi-particle quantum walk search can therefore be particularly useful on quantum simulator platforms where nearest-neighbor locality is a constraint and quantum walk algorithms are thus more natural to implement than gate-based Grover search.

Beyond answering our research question, we formally generalized the local interaction framework of Mittal and Sowi\'{n}ski~\cite{MitSo2026} to $P$ walkers and, more importantly, introduced what is -- to the best of our knowledge -- the first multi-particle quantum walk search algorithm. On the numerical side, we contributed to distributed simulations of quantum walks through the halo exchange algorithm of Sec.~\ref{sec:halo_exchange}. Here, we partitioned only the lattice of a single walker, but the implementation can be generalized to partition the lattices of several walkers, limited only by the computing resources available.

Regarding future work, we note that although we introduced two local interaction models, we did not attempt to engineer them to optimize the search itself. An interaction model might exist that drives stronger localization at the marked vertex and thereby improves the asymptotic scaling. Such an interaction could be either local or long-range, the latter of which we have not studied here. Furthermore, although amplitude amplification can already be applied after the multi-particle quantum walk search, Tulsi's algorithm reaches a constant success probability without the extra $\sqrt{\log N}$ overhead in the runtime~\cite{Tul2008}. We therefore suggest adapting it to the multi-particle setting.


\begin{acknowledgments}
We acknowledge fruitful discussions with Marcus Ka\-nes\-tr\o{m}, Eli\v{s}ka Greplová and Anton Akhmerov. D. O. O. acknowledges support from the Kavli Foundation. The authors acknowledge the use of computational resources from the DelftBlue supercomputer, provided by the Delft High Performance Computing Centre~\cite{DHPC2024}. 

{\it AI usage disclosure}
We have used Anthropic Claude Opus and OpenAI GPT models to setup several parts of the numerical simulations and plotting. All AI-generated code was carefully reviewed and tested by a human. 
\end{acknowledgments}

\begin{center}
    {\bf Data availability statement}
\end{center}
The code, generated data and plots for all simulations performed in this paper
are available on Zenodo and the source code is also maintained on
GitLab~\cite{MeOri2026}.

\bibliography{grover_bib}

\end{document}